\documentclass[letter]{aa}

\usepackage[varg]{txfonts}
\usepackage{natbib}
\usepackage{graphicx}
\usepackage{rotating}

\newcommand{\Msun}{M$_{\odot}$}

\begin{document}

\title{A  molecular outflow driven by the brown dwarf binary FU Tau \thanks{Based on observations carried out with the IRAM 30m Telescope. 
IRAM is supported by INSU/CNRS (France), MPG (Germany) and IGN (Spain).}}
\titlerunning{FU Tau outflow} 
\authorrunning{Monin et al.}

\author{J.-L. Monin \inst{1} \and E.T. Whelan \inst{2}  \and 
  B. Lefloch \inst{1} \and C. Dougados \inst{1} \and C. Alves de Oliveira \inst{3}}

\institute{UJF-Grenoble / CNRS-INSU, Institut de Plan\'etologie et d'Astrophysique de Grenoble (IPAG) UMR 5274, Grenoble, 38041, France
  \and Institut f\"{u}r Astronomie und Astrophysik, Kepler Center for Astro and Particle Physics, Sand 1, Eberhard Karls Universit\"{a}t, 72076 T\"{u}bingen, Germany, 
  \and European Space Astronomy Centre (ESA), P.O. Box 78, 28691 Villanueva de la Ca\~{n}ada, Madrid, Spain} 

\date{}

\abstract
{We report the detection of a molecular outflow driven by the brown dwarf binary FU Tau. Using the IRAM 30~m telescope we observed the $^{12}$CO(2-1) (CO) emission in the vicinity of FU Tau and detected a bipolar outflow by examining the wings of the CO(2-1) line as we moved away from the source position. An integrated map of the wing emission between 3~kms$^{-1}$ and 5~kms$^{-1}$  reveals a blue-shifted lobe at a position of $\sim$ 20~\arcsec\ from the FU Tau system and at a position angle of $\sim$ 20$^{\circ}$. The beam size of the observations is $11\arcsec$\ hence it is not possible to distinguish between the two components of the FU Tau binary. However as optical forbidden emission, a strong tracer of the shocks caused by outflow activity, has been detected in the spectrum of FU Tau A we assume this component to be the driving source of the molecular outflow. \\
We estimate the mass and mass outflow rate of the outflow at 4 $\times$ 10$^{-6}$~\Msun\ and 6 $\times$ 10$^{-10}$~\Msun/yr respectively. These results agree well with previous estimates for BD molecular outflows.  FU Tau A is now the third BD found to be associated with molecular outflow activity and this discovery adds to the already extensive list of the interesting properties of FU Tau. }

 
\maketitle 

\section{Introduction} 

Young brown dwarfs (BDs) occupy the mass regime between stars and planets and are therefore significant to any theory describing activity in star forming regions. Thus they have become the subjects of increased scrutiny in recent years \citep{Luhman2012}. Their formation mechanism is at present much debated and indeed it has been postulated that they may form by more than one mechanism \citep{Whitworth06}. 
The simplest idea is that they form in the same manner as low mass stars i.e. through the gravitational collapse of substellar mass cores \citep{Padoan04}. These cores occur directly by the process of turbulent fragmentation. In this scenario, BDs are just scaled-down versions of low mass stars.
Detailed studies of the circumstellar environments of young BDs provide critical constraints to different formation mechanisms and are needed to identify the dominant mechanism. In particular, if BDs form like low mass stars we expect their accretion/outflow properties to be analogous. As a low mass star forms it displays a series of ubiquitous observational properties, for example accretion disks, outflows, excess emission in the near-infrared and visual absorption. 

The observational evidences gathered to date in various wavelength domains indicate that young BDs show accretion and ejection behavior similar to low mass stars. For example they demonstrate T Tauri-like accretion \citep{Jay03, Natta2004, Monin2010, Rigliaco11} and both optical and molecular outflows, driven by BDs, have been detected. ISO-Oph~102 is a good example. It is an accretor with an observed accretion disk \citep{Natta02, Natta2004} and recent ALMA observations have detected millimetre sized grains in its disk \citep{Ricci12}. Its optical jet was discovered by \citet{Whelan05} through the spectro-astrometric analysis of the [O]$\lambda$6300 emission line.  Forbidden emission lines (FELs) like [OI]$\lambda$6300 are important coolants in shocks and therefore good tracers of jets. Traditionally jets from classical T Tauri stars (CTTSs) are investigated by studying their FEL regions. \cite{Phan2008} also detected a CO molecular outflow driven by ISO-Oph~102. The orientation of the blue and red lobes agreed with the optical observations. 

The question of outflow activity in BDs is an important one, as a sufficiently efficient outflow activity could provide an explanation as to why the central object mass  does not reach the H burning limit \citep{Bacciotti11, Whelan09b, Machida09}. 
Molecular outflows are an important large-scale expression of jet launching. Indeed, molecular outflows were one of the first observational manifestations of this process to be studied \citep{Reipurth01b, Bach96}. While giant Herbig-Haro (HH) flows are optically visible and composed of many HH objects, each group representing different episodes of mass ejection, molecular outflows begin when the powerful bipolar jets accelerate and drive outwards the molecular gas in the vicinity of their parent star. Although it is accepted that they are powered by the primary jet from the protostar, the exact way in which the jet interacts with the molecular material is still uncertain \citep{Cabrit97, Downes99, Downes07}. Molecular outflows are primarily detected in the CO molecule and thus millimeter observations have dominated the search for them. These outflows are mainly detected from Class 0 and I low mass stars which are still embedded in their natal material. Observations of molecular outflows driven by the more evolved Class II CTTSs are much rarer \citep{Cabrit11}. 

As of today, only two detections of molecular outflows from optically visible young BDs have been made so far \citep{Phan2008, PhanBao2011}, although it is postulated that due to the colder environment of BDs, molecular outflows may be more common than in CTTS. We have conducted a  survey of young BDs with the IRAM 30m telescope in the $^{12}$CO(1-0) and $^{12}$CO(2-1) to test this hypothesis (Whelan et al. 2013, in prep.). Their approach is to target BDs known to be accreting and which also show evidence of outflow activity primarily in form of FELs, in a mass range of 0.02~\Msun\ to 0.13~\Msun\, including a few very low mass stars (VLMSs). In this letter we report the detection of a remarkable molecular outflow in FU~Tau, as part of our IRAM survey.

FU Tau (04$^{h}$23$^{m}$35$^{s}$4, +25$^{\circ}$03$^{\arcmin}$03$^{\arcsec}$05) is a BD-BD binary with a projected angular separation of 5\farcs7 or 800~AU at the distance to Taurus and a position angle (PA) of $\sim$ 145$^{\circ}$ \citep{Luhman09}. Its membership of the Taurus molecular cloud has been known for some time \citep{Jones79} and it is situated in a relatively isolated region of the cloud. \citet{Luhman09} give the spectral type of FU Tau A at M7.25, corresponding to a mass of 50~M$_{\rm Jup}$ and the spectral type and mass of the companion at M9.5 and 15~M$_{\rm Jup}$ respectively. The wide nature of the FU Tau binary challenges models which suggest that BDs form when their accretion is halted due to ejection from their natal clouds, since the system appears to have formed irrespectively of dynamical interaction with nearby stars. 
A further intriguing property of FU Tau A is its over-luminosity with respect to other members of the Taurus star-forming region of the same spectral type \citep{Luhman09, Scholz11}. The spectral energy distributions (SEDs) of both components show excess emission indicating the presence of circumstellar disks and their disks are classified as being Class II by \citet{Luhman09}. Furthermore, optical spectra clearly show that accretion is on-going in FU Tau A. \citet{Stelzer10} estimated the mass accretion rate from both the H$\alpha$ and He I ($\lambda$5876) lines with $\dot{M}_{H\alpha}$ = 3.5 $\times$ 10$^{-10}$ M$_{\sun}$~yr$^{-1}$ and  $\dot{M}_{HeI}$ = 7.5 $\times$ 10$^{-10}$ M$_{\sun}$~yr$^{-1}$. Evidence of outflow activity comes from the detection of the [O I]$\lambda$ 5577 and [O I]$\lambda$ 6300 forbidden lines in an optical spectrum of FU Tau A \citep{Stelzer10}.
All of these facts combined show that the FU Tau system is probably a rarity amongst BDs and thus it is of considerable interest to test models describing the formation and evolution of BDs. For this reason we chose to publish the discovery of its molecular outflow separately from the global presentation of our IRAM survey. 

\section{Observations and Data Reduction}
\label{observations}

%


Observations of the CO(2-1)
line emission were carried out at the
IRAM 30m telescope on July 16th-18th 2011 using the EMIR receivers at 1.3mm. 

\begin{figure}[htb]
{\includegraphics[width = 7cm]{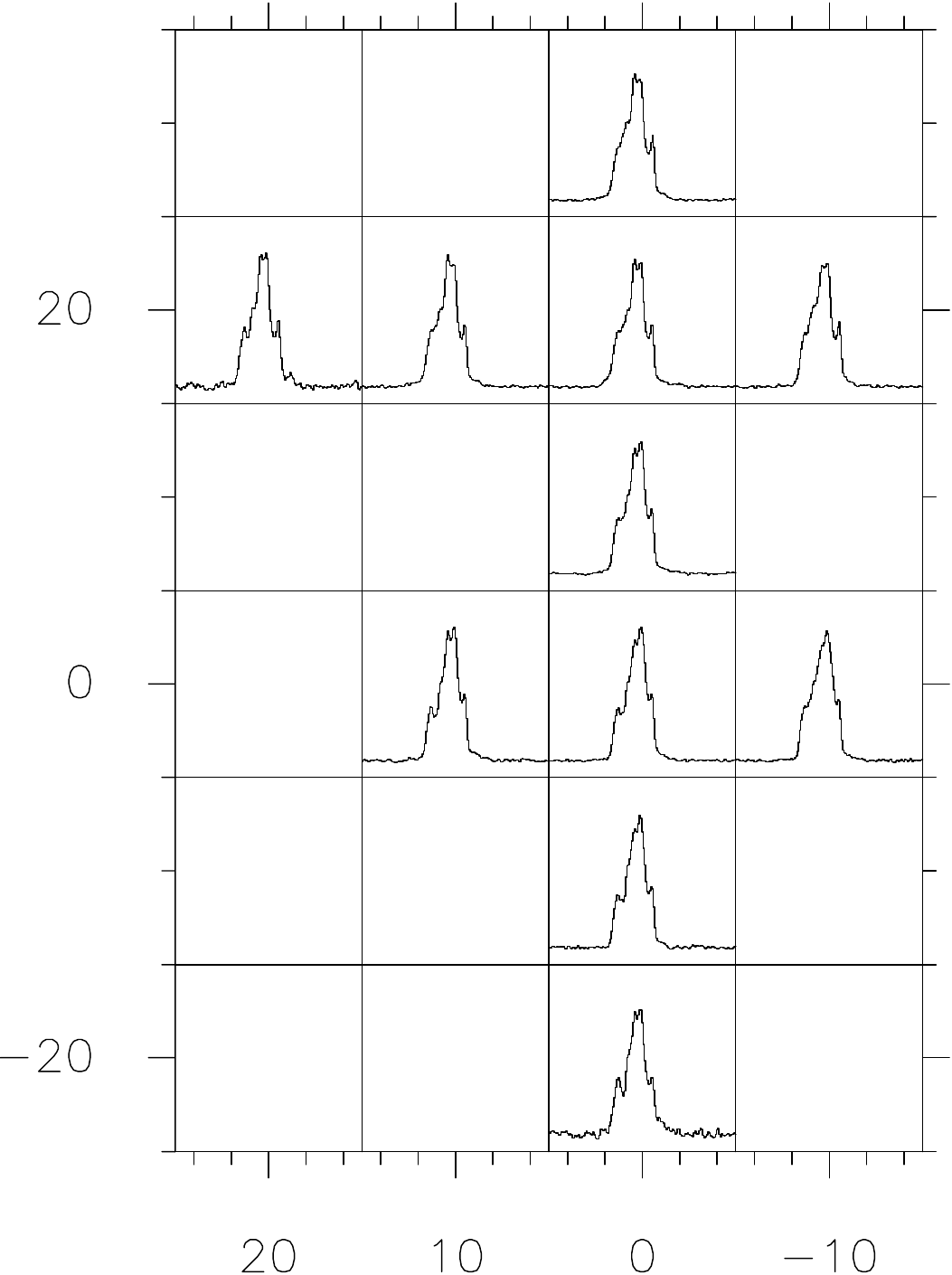}}
\caption{A Map showing the positions of the 11 spectra obtained for FU Tau. The CO(2-1) line is shown here. The scale of the grid is 0-15 kms$^{-1}$ and -0.6-6~K in x and y respectively. For  all spectra except the (0,-20) and (+20,+20) positions, the rms is less than 40 mK. For the (0,-20) and (+20,+20) positions  the noise is $\sim$ 80 mK. \label{fig:futau-map}}
\end{figure}

In a first step, deep integrations were performed towards the protostar and at
a reference position located $20\arcsec$ away. In a second step, in case of
significant variations of the CO emission between both positions, i.e. beyond
the $3\sigma$ intensity level, more extended mapping at $12\arcsec$ sampling
was performed. The CO emission map detected toward FU~Tau is displayed in Fig~\ref{fig:futau-map}.

Observations were carried out in Frequency Switch mode using a throw of
14.3~MHz at 1.3mm, with a phase time of 0.2 second.  An
autocorrelator providing us with a spectral resolution of 40 kHz was used as
spectrometer. The weather conditions were rather good and stable, with system
temperatures $T_{\rm sys}$ varying between 200 and 400\,K. Each position was
observed so to reach a final rms of about 40~mK unless exception per velocity interval of 0.1\.km.s$^{-1}$, 
after averaging both polarizations. 

Pointing was checked every 1.5 to 2 hours and was found to be very stable, with
pointing offsets corrections less than $3\arcsec$. The telescope parameters are
adopted from the IRAM webpage. At the frequency of the CO(2-1) line,
the main-beam efficiency of the telescope is 0.59  and the half-power
beamwidth is $11\arcsec$. The intensities of the measurements are expressed in units 
of main-beam brightness temperature $T_{\rm mb}$.

The data were reduced using the {\em Continuum and Line Analysis Single dish Software} (CLASS, a GILDAS software\footnote{http://www.iram.fr/IRAMFR/GILDAS/}). In some of the sources, the CO
mesospheric emission line  was detected close to the cloud emission, which
peaks at $v_{lsr}\simeq +6\,$km.s$^{-1}$, on the red side of the spectrum. For all our
observations we have adjusted a gaussian to the CO mesospheric line profile and
subtracted it out. The CO mesopheric line profile is typically a few K bright,
with a linewith of about $1\,$km.s$^{-1}$ (CHECK), much less than the velocity range of
the cloud emission and the outflow wing emission. When the outflow feature is
on the blueshifted part of the line spectrum, the mesospheric CO is absolutely harmless. When the observed outflow
wing is on the redshifted side, we checked that the CO mesospheric line is much
narrower than the wing velocity range, hence does not hamper the detection of
the latter.

\begin{figure}
\centering
{\includegraphics[width=6 cm, trim= 0cm 0cm 2cm 0cm clip=true, angle=0]{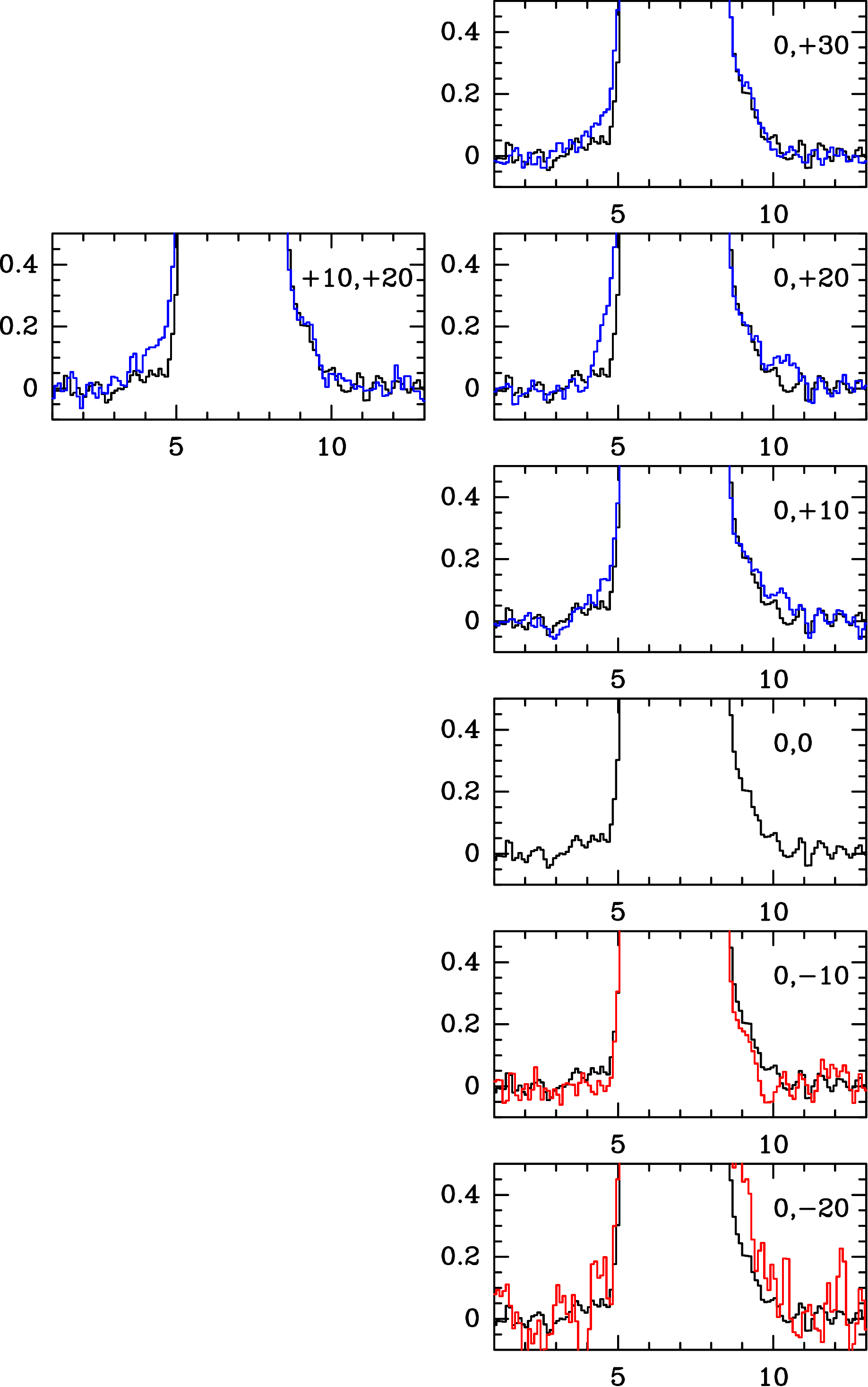}}
\caption{Montage of some of the CO(2-1) emission line profiles observed for FU Tau shown with a zoom on the region of the wings. The central (0,0) position spectrum is repeated as a dark solid line in all the plots, superimposed on the lines observed at the other positions, in color. We detect excess emission in the blue-wing between 3~kms$^{-1}$ and 5~kms$^{-1}$, at  the (0, 20), (10, 20) and (0, 30) positions. This points to a molecular outflow driven by the FU Tau system.\label{fig:wings}}
\end{figure}

\section{Results and Discussion}
\label{results}
\subsection{Outflow signature}
In Fig.~\ref{fig:wings} a magnified view of the wings of the CO(2-1) emission line at each point in the map of FU Tau (Fig.~\ref{fig:futau-map}) is shown. The central (0,0) position spectrum is repeated as a dark solid line in all the plots, superimposed on the color lines observed at the other positions. As the beam is $\approx 11\arcsec$ wide it encompasses both FU Tau A and FU Tau B. The bulk of the CO(2-1) emission comes from the cloud and is centered on the cloud velocity at $6\,$km.s$^{-1}$, and we search for outflow signature from variations in the CO(2-1) wing emission with respect to the emission on the central source. The outflow emission which is shifted in velocity with respect to the cloud is much fainter than the cloud emission and therefore it will be detected in the wings of the CO(2-1) emission line.  Fig.~\ref{fig:wings} shows that a blue component develops in the wing as we move towards the north, with an excess wing emission seen at the (0, 20), (10, 20) and (0, 30) positions between a velocity of 3~kms$^{-1}$ and 5~kms$^{-1}$. There is also a hint of blue-shifted excess emission at the (0, 10) position, and a red wing appears in the 8-10~km.s$^{-1}$ range in the (0,-20) position spectrum.
The detection of this excess emission strongly points to a molecular outflow driven by FU Tau. In Fig.~\ref{fig:futau-area} we present an integrated intensity map of the blue-shifted wing emission in the velocity range 3~kms$^{-1}$ to 6~kms$^{-1}$. The black squares mark the positions at which data was collected and we have super-imposed an optical image (WFCAM/UKIRT) of the FU Tau binary taken from \cite{Luhman09}. The detection of the outflow in the form of a blue-shifted lobe towards the north-east is clear. Without further data, we estimate a PA of $\sim$ 20$^{\circ}$ for the outflow axis.  

\subsection{Outflow Parameters}
\begin{figure}[htb]
\includegraphics[width = 7cm]{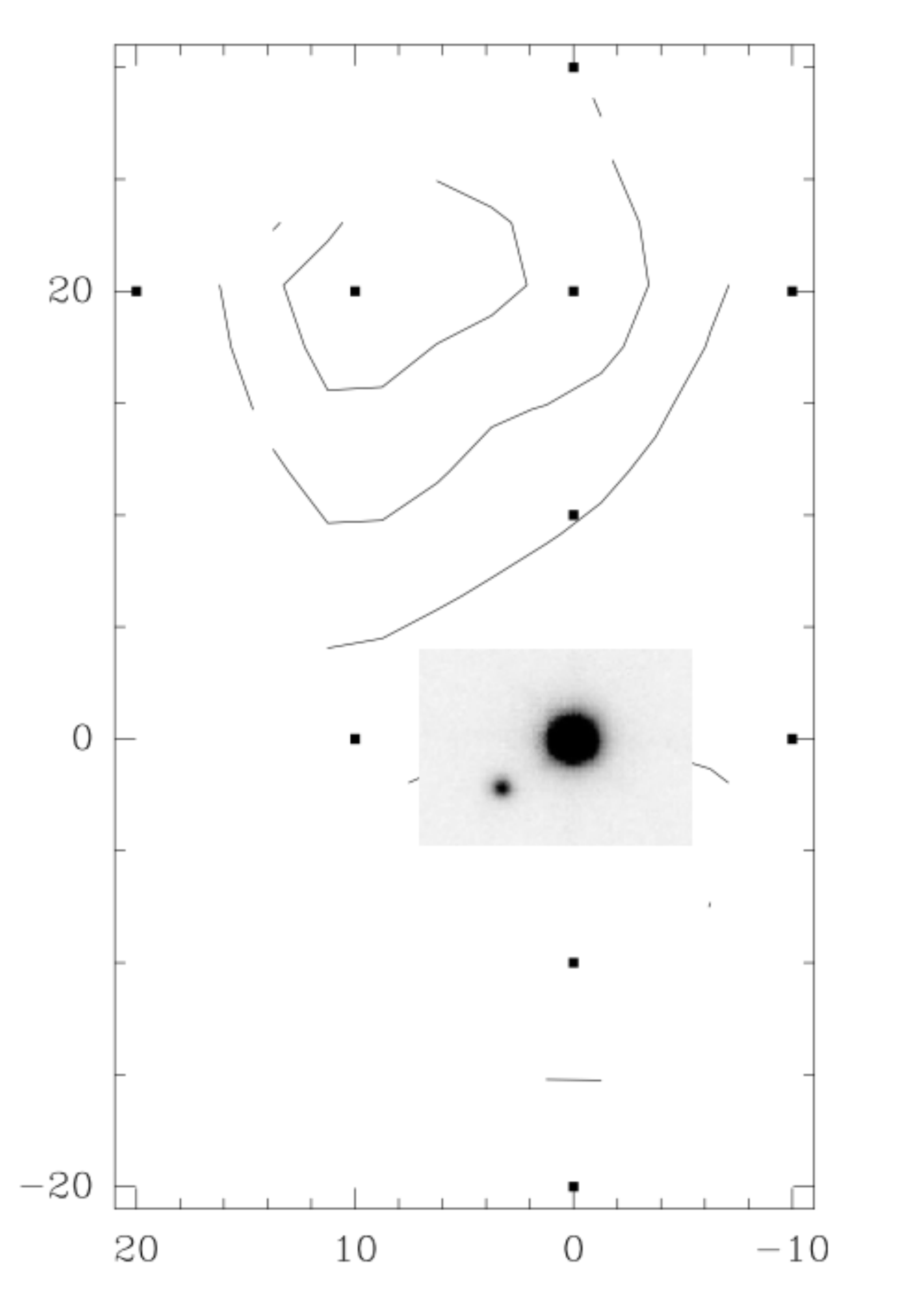}
\caption{An Integrated map of the blue-shifted wing emission in the velocity range 3-5 kms$^{-1}$. The LSR velocity of the BD is +6 kms$^{-1}$. The black squares mark the positions at which observations we made. The level are 0.1, 0.15, 0.2, and 0.25~K.km.s$^{-1}$. Clearly we see a blue outflow lobe at a PA of $\sim$ 20$^{\circ}$. We have superimposed a UKIDSS K-band image taken from \citet{Luhman09} of the BD binary at the same scale and assuming that FU Tau-A is at the central position.\label{fig:futau-area}}
\end{figure}

Following \citet{Bach90} we compute the CO column density in the blue lobe of the outflow with the following equation.
\begin{equation}({\rm CO})_{({\rm cm}^{-2})} = 1.06\,10^{13} \,T_{\rm mb} \exp({\frac{16.5}{T_{\rm mb}}}) \int T_R(2-1) \,dv\end{equation}
Adopting a gas excitation temperature  $T_{\rm mb} \approx 15\,$K, with an ${\rm H}_2/{\rm CO}$ ratio of $10^4$ and the results of Figure~\ref{fig:futau-area}, we compute a mass in the blue lobe of the flow of $M_B ({\rm H}_2) = 4\pm 0.8 \, 10^{-6}\,M_\odot$. 

If we suppose that the momentum of the underlying jet has entirely been transferred to the molecular component that we observe today, we can write : 
\begin{equation}M_B({\rm H}_2) <\!V_{\rm max}\!> \,= \dot{M}_{\rm jet} <\!V_{\rm jet}\!> \tau_{\rm dyn}\end{equation}
We measure $<\!V_{\rm max}\!> = 3\,$kms$^{-1}$, and we take a canonical value $<\!V_{\rm jet}\!> = 100\,$kms$^{-1}$ ; together with $\tau_{\rm dyn} \approx 200\,$yr (see \S~\ref{subsec:outpowsour}, first paragraph),  we obtain a mass loss rate for the blue-shifted lobe of  $\dot{M}_{\rm out} = 6\pm 1.3\, 10^{-10}\,M_\odot$/yr. This value can be modified by various factors. 
For instance, we can adopt a correction factor to take into account the fact that the jet might have been episodic in the past. \citep{PhanBao2011} use a factor of 10 for this purpose. Also, the excitation temperature is uncertain although the $T_{\rm mb}\exp(16.5/T_{\rm mb})$ factor does not vary much over $T_{21} = 10-25\,$K range. We could also take into account extinction effects and the fact that we are only measuring half of the flow emission. Thus this value must be taken as a first order estimation of the outflow rate and most probably underestimates the rate of the underlying jet. 
The outflow parameters are summarized in Table~\ref{tab:outflow-param}.

\begin{table}
\begin{center}
\begin{tabular}{lllll}
PA      & i         & N(CO)            & M(H$_2$) & $\dot{M}({\rm H}_2)$ \\
($^o$)   & ($^o$) & (cm$^{-2}$)     & ($M_\odot$) & ($M_\odot$/yr) \\
20      &  60     & $3.6\,10^{16}$ & $4\pm 1\,10^{-6}$ & $5\pm 1\,10^{-10}$
\end{tabular}
\caption{FU~Tau outflow parameters\label{tab:outflow-param}} 
\end{center}
\end{table}

\subsection{Outflow powering source}
\label{subsec:outpowsour}
Although we cannot disentangle FU Tau A from FU Tau B it is most likely that the outflow is driven by the primary as forbidden emission associated with the primary has already been detected \citep{Stelzer10}. Thus for the rest of the discussion we will assume that FU Tau A is the driver of the outflow. 
The peak of the blue lobe is measured at $\sim$ $20''$ from the central source, projected on the plane of the sky. Adopting a projection angle of 60$^o$ 
\citep{Stelzer2013}, 
the linear distance is $\approx 50''$, corresponding to $7000\,$AU at the distance of Taurus (140~pc). 
 At $100\,$kms$^{-1}$, this yields a dynamical age  $\tau_{\rm dyn}\approx\,200\,$yrs for the observed outflow event.

Previous to the results presented here, ISO-Oph~102 and MHO 5 with masses of 60~M$_{Jup}$ and 90~M$_{Jup}$ were the lowest mass objects for which molecular outflows were detected \citep{Phan2008, PhanBao2011}. The outflow mass and mass outflow rate were estimated at M$_{\rm out}$ = 1.6 $\times$ 10$^{-4}$ \Msun ; $\dot{M}_{\rm out}$ = 1.4 $\times$ 10$^{-9}$ \Msun/yr and M$_{\rm out}$ = 7.0 $\times$ 10$^{-5}$ \Msun ; $\dot{M}_{\rm out}$ = 9.0 $\times$ 10$^{-10}$ \Msun/yr respectively. Thus our estimates of the mass and mass outflow rate of the FU Tau outflow agree with previous results and are in line with the fact that FU Tau A has the smallest mass of the three objects. The derived values of  $\dot{M}_{\rm out}$ are also consistent with $\dot{M}_{\rm out}$ measured for the optical components of BD outflows \citep{Whelan09b}. For ISO-Oph~102 the outflow rate in the molecular component was found to be slightly higher than the the optical component. However, it is reasonable that $\dot{M}_{\rm out}$ for a molecular outflow could be greater than the outflow rate in the underlying jet. Assuming that the jet is powering the molecular outflow \citep{Downes07}, the mass outflow rate of the molecular component will grow with time as the jet transfers increasing amounts of energy and momentum. The size of the outflow of $\sim$ 20~\arcsec\ is compatible with the ISO-Oph~102 and MHO 5 flows and with observations of BD optical outflows. The agreement between the scale of the molecular and optical components is important if one is to accept that the jet drives the molecular flow. Finally we compare the mass outflow rate of the CO outflow from FU Tau with the derived mass accretion rate and find the ratio of mass outflow to mass accretion, $\dot{M}_{\rm out}$ / $\dot{M}_{\rm acc}$ = 0.8 to 1.7. $\dot{M}_{\rm out}$ / $\dot{M}_{\rm acc}$ measured for other BDs and VLMSs has also been found to be high compared to T Tauri stars where it is measured at 1-10\%  \citep{Bacciotti11, Whelan09b}. 
We stress that such a high value of $\dot{M}_{\rm out}$ / $\dot{M}_{\rm acc}$  cannot be used as a clue that this outflow removes or has removed a large fraction of the central object' mass (as in the \citet{Machida09} model), because the observed FU~Tau molecular outflow results from the entrainment by an underlying jet, and concerns an ejected mass several orders of magnitude smaller than in the \citep{Machida09} model. On the same hand, another explanation for the for $\dot{M}_{\rm out} / \dot{M}_{\rm acc}~{\rm ratio}~\approx 1$ in the current known series of BD sources could be due to an observational bias, because the first currently available observations are only sensitive to the most extreme jets in brown dwarfs. If this is true, the current ratios should prove much higher than the (as-yet unobserved) mean in BDs. More sensitive observations are thus needed to solve this issue.


\section{Summary}
The discovery of a molecular outflow driven by FU Tau A adds significantly to the interesting properties of this source and its binary companion. The FU Tau binary has a large separation compared to other binary systems and is thought to have formed in relative isolation. Both components harbor Class II accretion disks and FU Tau A is somewhat over-luminous for its spectral type. The fact that FU Tau A is driving an outflow demonstrates that despite having unusual characteristics it still exhibits properties which are strongly linked to the formation of low mass protostars. The mass, scale and mass outflow rate we measure for the FU Tau A CO outflow agrees with previous observations of BD molecular outflows.  While this result is a further important piece of evidence linking how BDs form to low mass star formation the derived ratio of mass outflow to accretion rates is much higher than what is observed in low mass protostars and in particular the T Tauri stars. For other BDs the two rates have been found to be comparable thus these new results for the FU Tau system support other studies of BD outflow activity \citep{Bacciotti11, Whelan09b}. $\dot{M}_{\rm out} / \dot{M}_{\rm acc}~{\rm ratio}~\approx 1$ in the current known series of BD sources could be due to an observational bias, because the first currently available observations could be only sensitive to the most extreme jets in brown dwarfs. More  observations are  needed to solve this issue. FU Tau is an excellent candidate for follow-up observations with sub-millimeter interferometer such as the Plateau de Bure interferometry or the Sub-miillimeter Array (SMA). With higher angular resolution observations we will be able to fully resolve the outflow, search for a red-shifted lobe and confirm whether FU Tau A is the driving source of the flow.


\begin{acknowledgements}The authors would like to acknowledge the help of the IRAM 30\,m team during the observations. E.T. Whelan is supported by an IRCSET-Marie Curie International Mobility Fellowship in Science, Engineering and Technology within the 7th European Community Framework Programme. We thank the referee, S. Mohanty, for a fast and thorough review that helped to improve the quality of this paper.
\end{acknowledgements}

\bibliographystyle{aa}
\bibliography{references}

%
%
%
%
%

\end{document}